\documentclass[12pt,letterpaper]{article}
\usepackage[a4paper, total={7in, 10in}]{geometry}

\usepackage{graphicx}
\usepackage{helvet}
\usepackage{authblk}
\usepackage{hyperref}
\usepackage{amsmath} 
\usepackage{amssymb} 
\usepackage{orcidlink} 
\usepackage[super,comma,sort&compress]  
   {natbib}
\usepackage[right]{lineno} \linenumbers
\usepackage[T1]{fontenc}

\makeatletter
\renewcommand{\maketitle}{\bgroup\setlength{\parindent}{0pt}
\begin{flushleft}
  \textbf{\@title}
  
  \@author
\end{flushleft}\egroup}
\makeatother

\title{Understanding the Dynamics of Evaporation-Driven Colloidal Self-Assembly}
\date{}

\author[1]{Junyu Yang}
\author[1]{Abhinav Naga}
\author[1]{Xitong Zhang}
\author[1,*]{Halim Kusumaatmaja}

\affil[1]{Institute for Multiscale Thermofluids, School of Engineering, The University of Edinburgh, EH9 3FD, United Kingdom}

\affil[*]{Correspondence: halim.kusumaatmaja@ed.ac.uk}

\begin{document}

\maketitle

\section*{SUMMARY}

Complex colloidal cluster morphologies are desirable for the fabrication of advanced materials, such as photonic crystals and meta-materials, and can be formed through evaporation-driven packing. By coupling lattice Boltzmann and discrete element methods, here we elucidate the rich interplay between fluid and particle dynamics during evaporation-driven self-assembly of spherical colloidal particles. We construct a regime diagram for a wide range of evaporation rates, interparticle friction coefficients, and particle numbers, identifying parameter regimes for open, closed, and minimal moment of inertia cluster configurations. Analyzing the competition between capillary, hydrodynamic, normal, and friction forces, we show that interparticle friction can exert a disproportionately strong influence on the final packing outcome despite being considerably smaller in magnitude than other forces at play. Our simulation results further highlight the potential for tuning colloidal cluster configurations via their dynamic trajectories. 

\section*{KEYWORDS}


colloidal clusters, self-assembly, surface tension phenomena, particle friction, lattice Boltzmann method

\section{INTRODUCTION}

The behavioral similarity between atoms and colloidal particles \cite{poon2004colloids, van2003chemistry} has inspired the application of molecular principles to the design of colloidal systems, particularly for bottom-up material fabrication via self-assembly. In this context, colloidal particles and clusters serve as fundamental building blocks for a wide range of functional materials \cite{hou2023interfacial, cai2021colloidal}. Self-assembly refers to the spontaneous organization of colloidal particles, driven either by specific interparticle interactions\cite{cui2021self, lyu2022low, chakraborty2022self},  external field manipulation \cite{zaibudeen2021dc, kang2022reconfiguring}, or mediated by environmental factors \cite{grzelczak2010directed, vialetto2024versatile}. To enable effective and controllable self-assembly, it is often desirable to employ anisotropic colloidal particles as building blocks\cite{rao2020leveraging, kim2021patchy}. A powerful strategy is to pre-assemble a finite number of colloidal spheres into well-defined colloidal clusters \cite{manoharan2003dense, meng2010free, hueckel2021total}, which in turn are assembled into designed superstructures through shape-complementary interactions \cite{avvisati2017fabrication, marson2019computational, beneduce2023engineering}. Consequently, the rational design and fabrication of target colloidal clusters is a critical step towards the realization of functional assemblies.

A widely adopted method for fabricating well-defined colloidal clusters is the evaporation-driven packing technique \cite{manoharan2003dense, wagner2010templated, zargartalebi2022self}. In this approach, a specific number $N$ of colloidal particles are encapsulated within emulsion droplets dispersed in surrounding fluids. As the emulsion evaporates, the particles are drawn together by capillary forces, forming compact clusters. Hitherto, both experimental and theoretical studies have focused on a regime of quasi-static, slow evaporation \cite{lauga2004evaporation, arkus2009minimal} and frictionless particle dynamics \cite{li2008forces, yethiraj2007tunable}. In this limit, such processes typically yield a structure for each $N$ that is characterized by a minimal moment of inertia of the mass distribution \cite{lauga2004evaporation, manoharan2003dense}. 

In practice, however, evaporation-driven self-assembly is a dynamic rather than a quasi-static phenomenon. The final packing structure results from a complex competition between capillary and hydrodynamic forces exerted by the liquid to the particles, as well as normal contact and friction forces between the particles. In recent years, there is strong interest to accelerate the droplet evaporation rate to speed up the self-assembly process. This often leads to poorly understood packing structures that deviate from those predicted by the minimal moment of inertia theory \cite{cho2007complex, zhou2022tuning}, and it motivates determining, as a function of evaporation rate, the yields of different packing configurations. Moreover, frictional interactions between particles are intrinsically complex and cannot be neglected~\cite{holtzman2012capillary,badetti2018shear,vivacqua2019analysis,comtet2017pairwise}. Recent studies in related contexts underscore the central role of particle friction. For example, particle friction can affect deposition patterns in drying droplets~\cite{xie2018dot}, self-cleaning on functional surfaces~\cite{naga2025modelingdropletparticleinteractionssolid}, and jamming transitions in granular systems~\cite{zhao2022ultrastable}. Accordingly, particle friction can also influence the resulting cluster morphology, including the requirement for cluster rigidity.

Taken together, we therefore identify two important, interrelated questions to address in this work. First, what are the upper limits on droplet evaporation rate and particle friction coefficient for which minimal moment packing remains achievable?
Second, can we engineer other, more diverse configurations by exploiting evaporation and friction dynamics? These require detailed understanding of the complex interplay between droplet and particle dynamics.

While experiments provide valuable insights, numerical simulations offer a more effective means of analyzing the forces and cluster rearrangements involved in self-assembly. Particle-based models such as molecular dynamics \cite{katiyar2019evaporation} and Monte Carlo simulations \cite{flavell2023programmed} have been used to study evaporation-driven packing. However, these approaches often ignore hydrodynamic and frictional effects, which are central to our investigation. To overcome this limitation, we employ a recently developed hybrid simulation framework that integrates the lattice Boltzmann (LB) method with the discrete element method (DEM) \cite{zhang2020new, naga2025modelingdropletparticleinteractionssolid}. This framework enables accurate modeling of the emulsion-assisted packing process by fully capturing capillary, hydrodynamic, normal contact, and frictional forces. Using this approach, we conduct a detailed analysis to reveal how evaporation dynamics and friction jointly determine the resulting packing configurations in colloidal clusters. We believe these results are valuable for developing rational design principles for controlled assembly of target colloidal structures. 

\section{RESULTS}

To systematically investigate the importance of droplet and particle dynamics, we incorporate all relevant forces into the simulation. These include capillary forces ($\textbf{F}_\text{c}$), hydrodynamic forces ($\textbf{F}_\text{h}$), normal contact forces ($\textbf{F}_\text{n}$), and frictional forces ($\textbf{F}_\text{f}$), as illustrated in Figure~1a.
For simplicity, here a uniform density $\rho$ is assigned to the particles, emulsion droplets, and surrounding fluid, and the same kinematic viscosity $\nu$ is applied to both fluids. This assumption reflects typical experimental systems, such as toluene–water emulsions containing polystyrene particles\cite{manoharan2003dense}. The particle–emulsion contact angle is fixed at 30$^\circ$, a representative value for lipophilic particles, ensuring that capillary forces drive the particles inward to form compact clusters.
More generally, since there is little convection during the self-assembly process, density and viscosity contrasts between the emulsion droplet and the surrounding fluid do not play significant roles on the packing dynamics. As demonstrated in Video~S1, the resulting structures remain the same even when the dynamic viscosity ratio $\rho_\mathrm{sur}\nu_\mathrm{sur}/(\rho_\mathrm{emul}\nu_\mathrm{emul})$ differs by up to a factor of 100 if we ensure the other emulsion properties and evaporation rate remain the same.

In this work, the droplet interface recedes at a constant evaporation rate $U_\text{e}$, and both interface evolution and particle trajectories are tracked throughout the simulation. A representative snapshot is shown in Figure~1b, replicating experimental observations \cite{manoharan2003dense}. The self-assembly process proceeds as follows.
(i) Initially, colloidal particles are randomly distributed inside a large droplet, where surface tension does not significantly constrain their motion.
(ii) As evaporation continues, the interface contracts at a constant rate $U_\text{e}$, and surface tension gradually brings the particles together. Once the droplet volume falls below a critical threshold, particle–particle contacts begin to form.
(iii) Further evaporation accentuates the role of capillary forces, promoting structural rearrangement towards more ordered configurations.
(iv) Eventually, the process stabilizes with the formation of a well-defined cluster.

\subsection{Packing regime}

A wide range of evaporation rates $U_\mathrm{e}$ and friction coefficients $\mu$ are considered to construct a regime diagram that captures the combined effects of evaporation dynamics and frictional interactions, as shown in Figure~2. In the simulations, the Reynolds number ${Re}=U_\mathrm{e}d/\nu$ is extremely low, ranging from $2\times10^{-3}$ to $2\times10^{-1}$, where $d$ is the particle diameter. Therefore, inertial effects are negligible compared to viscous and capillary forces. The Capillary number $Ca = \rho \nu U_\mathrm{e} / \gamma$ thus serves as an effective dimensionless number to characterize the influence of $U_\mathrm{e}$, where $\gamma$ is the interfacial tension. $Ca$ measures the balance between viscous resistance and capillary driving forces. As demonstrated in Video S2, simulations with identical $Ca$ result in the same final packing configuration. Moreover, we can record the evolution of the second moment $M_2$. The second moment is defined as $M_2 = \sum_{i=1}^{N} r_{ic}^2$, where $r_{ic}$ is the distance between particle $i$ and the cluster center. $M_2$ is proportional to the moment of inertia and thus quantifies the compactness of the configuration. For a given Capillary number $Ca$, as shown in Figure~S1, the trajectories of $M_2$ collapse onto each other even when the evaporation rate $U_\mathrm{e}$ is varied.

Because small clusters ($N\le 4$) admit a single closed-packed configuration~\cite{guzowski2015droplet}, we focus on numerical simulations with $N\ge 5$. Figure~2 shows a representative regime diagram for $N=6$. For each point in the diagram, we compile statistics from 50 realizations with distinct initial particle distributions (Figure~S2). Results for other particle numbers ($N=5,7,8$) are provided in Figures~S3–~S5. Across $N$, the regime diagrams share the same qualitative structure; differences arise mainly in the positions of the boundaries. For ease of comparison, the corresponding boundaries for different $N$ are superimposed as dashed lines in Figure~2(b).

We first discuss the effect of the evaporation rate $ U_\mathrm{e} $ on the packing structure of colloidal clusters, and for simplicity, let us initially consider smooth particles with negligible friction ($\mu < 0.05$). With varying $Ca$, we identified three distinct categories of final packing configurations: open, closed, and minimal moment packings, as illustrated in Figure~2a. The three regions in the regime diagram in Figure~2b are defined based on the occurrence probabilities of open, closed, and minimal moment packings. In the minimal moment regime (blue region in Figure~2b), the minimal moment packing is obtained with $100\%$ probability. In the closed regime (red region in Figure~2b), both closed and minimal moment packings can appear. In the open regime (green region in Figure~2b), all three types of structures may occur.

For large evaporation rates ($Ca > 0.05$; indicated in green in Figure~2b), hydrodynamic forces dominate, and capillary attraction is not sufficient to draw colloidal particles into close contact. In this regime, the final structures are typically open packings with some colloids having fewer than three contact points. These allow for loose aggregates with structural flexibility, such as chains or planar assemblies (Figure~2a), which are more strongly influenced by the initial spatial arrangement of particles than by packing dynamics. 

For slow evaporation rates ($Ca < 0.003$, indicated in blue in Figure~2b), capillary forces dominate and effectively drive particles into the minimal moment configuration with $100\%$ probability, regardless of the initial particle distribution. The minimal moment packing represents a special subclass of closed configurations characterized by the minimization of the moment of inertia. For $N = 6$, the minimal moment structure has $O_{h}$ symmetry\cite{ameta2016chemical}, see Figure~2a in blue. Such structures are frequently observed in experiments conducted under slow evaporation rates \cite{manoharan2003dense}, as well as in theoretical analyses where the packing process is assumed to be quasi-static \cite{lauga2004evaporation}. In this configuration, the total number of particle contacts $N_{\mathrm{c}}$ reaches $3N - 6$, thereby satisfying the rigidity criterion and resulting in a mechanically stable structure. 

For intermediate evaporation rates ($0.003 \le Ca \le 0.05$, indicated in red in Figure~2b), capillary and hydrodynamic forces are comparable in magnitude. Although capillary attraction promotes initial aggregation, viscous resistance can hinder the subsequent rearrangements needed to minimize the moment of inertia. Under these conditions, in addition to the minimal moment $O_h$ structure, other closed-packed configurations are observed, such as $C_{2v}$, as illustrated in Figure~2a. These structures are characterized by each particle having at least three contacts, rendering the cluster relatively stable, while the moment of inertia is not minimal. We refer to such configurations as closed packings.

As shown in Figure~2c, with increasing $Ca$, the $C_{2v}$ structure becomes increasingly likely: at $Ca=10^{-2}$ it occurs with a probability of $10\%$, rising to $48\%$ at $Ca=0.05$. The $D_{3h}$ configuration matches the structure observed experimentally under high evaporation rates~\cite{cho2007complex}. In addition to $C_{2v}$ and $D_{3h}$, we also identify other closed packings not reported experimentally as shown in red in Figure~2a.

As shown in Figures~S3–~S5, increasing the particle number $N$ enlarges the diversity of closed packings and renders the minimal moment configurations more intricate. Consequently, the formation of minimal moment packings becomes increasingly constrained at larger $N$. To quantify this trend, we measured the probability of non–minimal moment packings as a function of the capillary number ${Ca}$ (Figure~S6a). We define the ${Ca}$ value at which the probability of closed packings first becomes nonzero as the boundary between the closed and minimal moment regimes. Our results, see dashed lines in Figure~2b, show that a lower ${Ca}$ suffices for the emergence of non–minimal moment structures because the landscape of ground states becomes more complex at larger $N$. 

The boundary shift with increasing $N$ saturates and eventually plateaus for $N>8$ (Figure~S6b). To demonstrate this, we simulated larger clusters ($N=10$, $N=12$, and $N=15$) at ${Ca}=10^{-4}$; in all cases, the minimal-moment structures $D_{4d}$, $I_h$, and $D_{3h}$ formed with $100\%$ probability. These results suggest that, for $N>8$, the closed–minimal boundary asymptotically approaches ${Ca}=10^{-4}$, as shown in Figure~2b. It is worth noting that previous experimental~\cite{manoharan2003dense} and theoretical~\cite{lauga2004evaporation} studies have typically considered systems with a limited number of particles, $N \leq 15$. As $N$ increases further, the packing dynamics are expected to become more complex and may give rise to additional packing structures \cite{marin2023colloidal}. Owing to computational cost, we restrict the present study to particle numbers in this range, leaving a systematic exploration of larger $N$ for future work.

We have also explored the boundary between open and closed packings for different $N$. As shown in Figure~S6a, the occurrence probability of open packings remains zero for $Ca<0.05$ across all particle numbers. Consistently open packings are observed only when ${Ca}>0.05$, independent of $N$. Therefore, the boundary between open and closed packings remains fixed at ${Ca}=0.05$ for all particle numbers considered. 

As shown in Figure~2b, for smooth particles, the conditions required to form closed but not minimal moment packings can be limited. For instance, we can consider $N = 6$ for a representative toluene-in-water emulsion system \cite{manoharan2003dense}. With a dynamic viscosity of $\eta \approx 0.3\ \mathrm{mPa \cdot s}$ \cite{nistwebbook} and an interfacial tension of $\gamma \approx 30\ \mathrm{mN/m}$ \cite{bkak2016interfacial}, the transition values of $Ca \sim 0.003$ and $Ca \sim 0.05$ correspond to $U_\mathrm{e} \sim 0.3\ \mathrm{m/s}$ and $U_\mathrm{e} \sim 5\ \mathrm{m/s}$. Such high evaporation rates are difficult to achieve experimentally unless extremely high temperatures are applied. Importantly, this suggests that the formation of closed but not minimal moment packing, as observed experimentally \cite{cho2007complex}, means interparticle friction is non-negligible. Depending on the roughness of the colloidal particles, the particle friction coefficient can vary significantly. Even with particles that are considered frictionless, upon increasing the normal contact force, there can be a critical threshold for a transition from a low-friction to a high-friction regime\cite{comtet2017pairwise}. According to our numerical results, the magnitude of the normal contact force acting on each particle scales as $F_\mathrm{n} \sim \gamma d$. For a representative particle-oil-water system with $\gamma \approx 30\ \mathrm{mN/m}$ and $d = 20\ \mu\mathrm{m}$, the resulting contact force exceeds $100\ \mathrm{nN}$, which is well above the critical threshold of $10\ \mathrm{nN}$ beyond which frictional interactions become significant due to surface asperities \cite{comtet2017pairwise}. Under such conditions, colloidal particles typically exhibit friction coefficients in the range of 0.1 to 0.5~\cite{holtzman2012capillary, badetti2018shear, vivacqua2019analysis, comtet2017pairwise}. 

To understand the role of friction, we perform dynamic simulations of colloidal self-assembly across a range of particle friction coefficients, as shown in Figure~2b. For particles with $\mu>0.05$, the conditions required to enter the closed packing regime become progressively easier to meet as friction increases, because frictional resistance suppresses the rearrangements needed to attain minimal-moment states. As illustrated in Figure~2d and e, even when $Ca$ would produce minimal moment packings for smooth particles, increasing friction leads to additional closed structures—including $C_{2v}$, $D_{3h}$, and other configurations with $N_{\mathrm c}<3N-6$. 

Based on our simulation results, the transition boundary between closed and minimal moment packings for $\mu>0.05$ follows an apparent power-law relation:
\begin{equation}
Ca = 2.5 \times 10^{-7} \mu^{-3.4}.
\label{powerlaw}
\end{equation}
This relationship highlights the high sensitivity of packing behavior to friction. A change of one order of magnitude in $\mu$ corresponds to a 3.4 orders of magnitude change in $Ca$. For example, when $\mu = 0.5$, the required $Ca$ to achieve closed packing reduces to $5.0 \times 10^{-6}$, which corresponds to an evaporation rate of $U_\mathrm{e} = 5 \times 10^{-4}\ \mathrm{m/s}$ in a typical toluene-water system. These results suggest that accounting for interparticle friction substantially broadens the range of conditions under which closed packings become relevant experimentally. Across particle numbers $N$, the boundary between closed and minimal moment packings remains unchanged, continuing to follow the power-law relation in Equation~\eqref{powerlaw} (see Figures~S3–~S5). This insensitivity to $N$ underscores the important role that friction can play in determining the packing outcome. 

Taken together, the regime diagram provides useful guidance for the experimental design of colloidal clusters. It suggests we can select desired structures by tuning the evaporation dynamics and interparticle friction.

\subsection{Force analysis}

To elucidate the dynamics underlying the regime diagram, we analyze the forces acting on particles during evaporation-driven packing. These are valuable information that will be difficult to obtain in experiments. First, to understand why higher evaporation rates can yield closed packings with non-minimal moments, we examine smooth particles under two representative conditions for $N=6$: ${Ca}=1.0\times10^{-3}$ and ${Ca}=0.05$, which—starting from identical initial particle distributions—produce the $O_h$ (minimal moment) and $C_{2v}$ (closed) structures, respectively. The only difference between these two cases is the Capillary number ${Ca}$. Figure~3a and b show the temporal evolution of the force components acting on a representative particle (marked in red). Forces are reported in non-dimensional form, $F^{*}=F/(\pi d \gamma)$.

As shown in Figure~3a and b, the evaporation-driven packing process separates into two successive stages: an initial packing stage (green) followed by a rearrangement stage (purple), with the transition defined by the onset of critical packing. We define critical packing as the instant when particles first become compacted and normal contact forces develop on each particle. By examining results from different simulations, we find that critical packing systematically occurs when the total number of contacts reaches $N_c = 2N - 4$. This value is two-thirds of the rigidity threshold $N_c = 3N - 6$ and corresponds to an average of two contacts per particle, indicating that, due to droplet shrinkage, most particles have already come into mutual contact. This corresponds to the condition for the formation of critical packing.

By comparing Figure~3a and b, pronounced differences already emerge at the initial stage. At the lower evaporation rate ($Ca=1.0\times10^{-3}$; Figure~3a), capillary forces are effectively the only active contribution, with other forces nearly zero; capillarity therefore dominates and governs particle motion. During this stage, the droplet remains nearly spherical until reaching its critical packing volume at $t^*=0.25$, and interfacial tension strongly constrains particle trajectories. The resulting cluster thus forms a highly symmetric configuration, which naturally relaxes into the energetically favorable $O_h$ structure during the subsequent rearrangement stage. In contrast, at the higher evaporation rate (Figure~3b), capillary and hydrodynamic forces are comparable in magnitude ($F^*\sim 0.2$) during the initial packing stage. The droplet shape departs from sphericity, 
and the critical packing formed at the end of this stage exhibits reduced symmetry and compactness. During the subsequent rearrangement stage, this asymmetric intermediate facilitates further reorganization into a $C_{2v}$ cluster. For the other particles, the force profiles are qualitatively similar (Figure~S7). 

Our simulation insights suggest that, for smooth particles within the closed packing regime, controlling the structure at critical packing enables control over the yield of the target packing. To test this, we tracked the evolution of the total contact number $N_{\mathrm c}$ and the cluster second moment $M_2$ over 50 realizations with distinct initial particle distributions for $N=6$ (Figure~3c and d).

Figure~3c and d compare typical minimal moment ($Ca=10^{-3}$) and closed ($Ca=0.05$) conditions (in the phase diagram) for smooth particles. Initially, when $N_\mathrm{c}<4$, the particle distribution is effectively random (e.g., the blue and red trajectories in Figure~3d are intertwined). As contacts begin to form ($N_\mathrm{c} \ge4$), at $Ca=10^{-3}$ (Figure~3c) the capillary force dominates the hydrodynamic force, and all the $M_2$–$N_\mathrm{c}$ trajectories converge to a single point corresponding to the $O_h$ structure ($M_2=12r^2$). 

In contrast, for $Ca=0.05$, hydrodynamic resistance becomes significant and diminishes the influence of capillarity. A subset of trajectories converges to the point associated with the $C_{2v}$ structure ($M_2=13.41r^2$), while another subset to $O_h$. The two families of trajectories exhibit a clear separation
(Figure~3d), especially after the critical packing condition of $N_\mathrm{c} = 2N-4 = 8$. We fit a polynomial to the lower envelope of the $C_{2v}$ group, yielding the dividing curve shown in black:
\begin{equation}
M_2/r^2=f(N_\mathrm{c})=0.06(N_\mathrm{c}-(3N-6))^2+M_2(C_{2v})/r^2.
\label{M2Nc}
\end{equation}
Equation~\eqref{M2Nc} also applies to other particle numbers. For example, Figure~S8 shows the case for $N=8$. These observations suggest a strategy for controlling packing yield: within the closed packing regime, increasing the cluster second moment $M_2$ at the critical-packing stage enhances the yield of non-minimal moment configurations.

These findings provide practical guidance for producing specific packing structures in experiments. Minimal moment packings are relatively easy to obtain by simply reducing the evaporation rate as much as possible (e.g. $Ca<10^{-3}$). In contrast, to generate non–minimal moment structures such as $C_{2v}$, we firstly should use a relatively high evaporation rate such that the packing dynamics operate within the closed regime (e.g. $Ca>10^{-2}$). Secondly, the detailed distribution of particles on the interface is crucial. Distributing particles at the droplet interface is useful to promote the formation of $C_{2v}$. Such configuration is commonly observed in Pickering emulsions\cite{ni2022pickering}.

To further illustrate the effect of initial particle distribution, we consider the two following setups: in one case, particles are concentrated on one side of the droplet; in the other, they are dispersed along the droplet equator. Numerical simulations, see Figure~4, show that, for sufficiently high evaporation rates, the yield of the $C_{2v}$ structure reaches $100\%$ when particles are dispersed along the equator. This suggests that achieving high yields of non–minimal moment structures experimentally can be explored by tuning the initial dispersion of particles on the droplet interface, for example by introducing external constraints or particle-particle interactions such as electrostatic repulsion \cite{bianchi2013self, li2024evaporative}.

We also find that varying the initial particle concentration does not affect the final packing structure distribution. By changing the initial droplet radius $R$, we simulated packing for three different initial concentrations (for the same total number of particles), as shown in Figure~S9. Under the conditions $Ca = 0.05$ and $\mu = 0.01$, the occurrence probability of the $C_{2v}$ structure is $50\%$ in all three cases. Thus, the initial particle concentration does not determine the yield of the final packing.

The regime diagram in Figure~2b, together with the power-law boundary in Equation~\eqref{powerlaw}, indicates that friction has a disproportionately influence on the final packing structure, largely independent of the particle number $N$. To elucidate the underlying mechanism, we perform a particle-resolved force analysis throughout the packing process. Figure~5 shows the temporal evolution of the force components acting on a representative particle (marked in red) for $N=8$ under two contrasting conditions that share the same initial particle distribution: minimal moment packing (${Ca}=1.0\times10^{-3},\ \mu=0.01$) and closed packing (${Ca}=1.0\times10^{-3},\ \mu=1.0$). The only parameter that differs between these two cases is the particle friction coefficient $\mu$.

During the initial packing stage, particles are freely distributed within the droplet and experience negligible forces. As evaporation shrinks the droplet, the system approaches critical packing at $t^*=0.5$. Up to this point, particle motions and force signatures are indistinguishable under both conditions, yielding identical critical packing configurations (see the bottom panels of Figure~5a and b). Differences arise only during the subsequent rearrangement stage.

In the minimal moment packing scenario (Figure~5a), the capillary force undergoes a rise-and-fall process, reaching its peak around $t^* = 0.75$. Between $t^* = 0.5$ and $0.75$, the shrinking droplet radius leads to a gradual increase in capillary force. This capillary force acts as the driving force that promotes the reorganization of particles from the critical packing state toward the minimal moment configuration. The normal contact force provides mutual constraints between particles.
This results in the formation of the minimal moment packing at $t^* = 0.75$. As evaporation continues, the droplet shrinks further, reducing the contact area between the particles and the droplet surface, which in turn weakens the capillary force. Eventually, the droplet fully evaporates.

In contrast, for rough particles (Figure~5b), the rearrangement follows a different pattern. Due to interparticle friction, relative motion between particles is hindered. As a result, when the capillary force reaches its peak at $t^* = 0.8$, the particles are not yet fully compacted, with the total number of contacts still less than the rigidity threshold ($N_\mathrm{c} < 3N - 6$). The frictional force creates a self-locking effect among the particles, preventing further rearrangement into the minimal moment configuration. Once the cluster becomes compact, relative motion between particles ceases, and dynamic friction transitions to static friction. Therefore, although the magnitude of the friction force is small compared to other forces, its effect becomes significant during the rearrangement stage.
 
Force profiles for the remaining particles are largely consistent (Figure~S10), confirming that friction governs the final structure during the rearrangement stage. For clarity, we provide Video~S3, which offers a detailed explanation of the packing dynamics described above.

These observations further support the non-linear relation between the capillary number $Ca$ and friction coefficient $\mu$ for the regime boundary in Equation~\eqref{powerlaw}. The pronounced contrast between evaporation-rate and friction effects captured by the power law arises from the fundamentally different ways in which hydrodynamic and frictional forces influence particle motion. As illustrated in Figure~S11a, hydrodynamic forces originate from the relative motion between particles and fluid, acting as a distributed viscous resistance that decelerates particles. This smooth drag can slow rearrangements and, if sufficiently strong, prevent clusters from reaching minimal moment packings before the droplet fully evaporates. By contrast, friction exerts a more direct and threshold-like resistance. Since it arises from solid–solid contacts rather than a continuous fluid phase, dynamic friction between particles transitions abruptly to static friction once the friction coefficient exceeds a critical value.

Figure~S11b shows the evolution of particle trajectories and interparticle friction during the packing process at the same $Ca=5\times10^{-3}$ but with different $\mu$. When the friction coefficient increases from $0.05$ to $0.1$, crossing the boundary in Figure~2, both the particle trajectories and the evolution of the friction force change markedly: the packing transitions from minimal moment to closed, and the friction regime shifts from dynamic to static friction. As the friction coefficient increases further ($\mu>0.1$), the force characteristics remain essentially unchanged. This indicates that the dominant effect of friction arises from the transition between dynamic and static friction. Consequently, a relatively small change in $\mu$ is sufficient to restore a jammed state even under appreciably slower evaporation. 

Since the key factor determining the final packing structure is whether the interparticle friction is in the dynamic or static regime, this transition provides insights into the power-law relationship in Equation~\eqref{powerlaw}. According to Equations (S19) and (S20), when energy dissipation is neglected, the transition is governed by the comparison between the tangential spring $\delta_\mathrm{t}$ and $\mu F_\mathrm{n}/k_\mathrm{t}$, with the threshold given by
\begin{equation}
\delta_\mathrm{t} = \mu F_\mathrm{n}/k_\mathrm{t}.
\label{threshold}
\end{equation}

We analyze the self-assembly trajectories under different $Ca$ conditions and track the evolution of $\delta_\mathrm{t}$, as shown in Figure~S12. Figures~S12a and b present the evolution of $\delta_\mathrm{t}$ for two representative cases near the boundary at $\mu=0.1$. Here, $\delta_\mathrm{t}$ is non-dimensionalized as $\delta_\mathrm{t}^* = k_\mathrm{t} \delta_\mathrm{t} / (\pi d \gamma)$ and compared with the dimensionless normal contact force ${F}_\mathrm{n}^*$. The most pronounced difference occurs in the interval $t^* = 0.6-0.8$. In this period, for the closed packing case at $Ca = 1\times10^{-3}$, $\delta_\mathrm{t}^*$ remains smaller than $\mu{F}_\mathrm{n}^*$, and the particles exhibit negligible relative motion. In contrast, when $Ca$ decreases to $5\times10^{-4}$, the capillary driving force becomes sufficiently strong to overcome the resistive forces, leading to significant tangential displacement between particles. As a result, $\delta_\mathrm{t}^* \simeq \mu{F}_\mathrm{n}^*$, sliding occurs, and the system evolves toward a minimal moment packing structure.

To extract the power-law relation, it is necessary to quantify the dependence of both $\delta_\mathrm{t}$ and $F_\mathrm{n}$ on $Ca$. Figure~S12c shows that ${F}_\mathrm{n}^*$ remains nearly unchanged across different $Ca$, with all curves collapsing onto the same order of magnitude. This is expected, since the normal force primarily balances the capillary driving force, and thus their magnitudes should remain comparable, as also confirmed in Figures~3 and 5. Therefore, the variation in frictional state is primarily controlled by $\delta_\mathrm{t}$. To quantify this behavior, we compute the time-averaged tangential spring $\overline{\delta_\mathrm{t}^*}$ and examine its dependence on $Ca$, as shown in Figure~S12d. In the closed packing regime, we find $\overline{\delta_\mathrm{t}^*} \sim Ca^{-0.3}$. Substituting this scaling into Equation~\eqref{threshold} yields $\mu \sim Ca^{-0.3}$ at the boundary between closed and minimal moment packings, which is consistent with the proposed power law in Equation~\eqref{powerlaw}. In contrast, in the minimal moment regime, where friction is dynamic, $\delta_\mathrm{t}$ is constrained by Equation~\eqref{threshold} and no longer increases with decreasing $Ca$. Thus, the variation of $Ca$ controls particle motion, which in turn determines $\delta_\mathrm{t}$ and the frictional state. Moreover, as shown in Figure~S12e, for $\mu=0.5$, $\delta_\mathrm{t}$ and $Ca$ follow the same power-law relation as for $\mu=0.1$ in Figure~S12d, further confirming that the power law in Equation~\eqref{powerlaw} holds across a wide range of friction coefficients.

To further substantiate the decisive role of friction during the rearrangement stage, we plot the evolution trajectories in the $M_2$–$N_{\mathrm c}$ plane with friction and compare them with Figure~3c and d (Figure~S13). The pairs in Figure~S13a,c and ~S13b,d share identical ${Ca}$ but differ in the friction coefficient $\mu$. Comparing Figure~S13a and ~S13c, the trajectories are nearly identical during the initial stage ($N_{\mathrm c}<8$) when ${Ca}$ is fixed; at critical packing, most cluster $M_2$ values lie below the dividing curve. The distinction arises in the rearrangement stage: due to friction, some clusters cannot converge further and their trajectories settle at $N_{\mathrm c}<3N-6$, yielding closed packings. A similar trend is observed at ${Ca}=0.05$ in Figure~S13b and ~S13d, where friction during rearrangement prevents clusters that would otherwise form $O_h$ or $C_{2v}$ from reaching $N_{\mathrm c}= 3N-6$.

From the above analysis, the yield of a given packing structure can, in practice, be controlled by tuning the interparticle friction coefficient. This coefficient depends on factors such as particle material, surface roughness, coating properties, and surface energy~\cite{butt2018surface, israelachvili2011intermolecular}. Experimentally, friction can be tuned by modifying surface roughness ~\cite{scherrer2025characterizing}, applying tailored surface coatings~\cite{lahiri2023low}, or adjusting the presence of adsorbed or chemically bound molecules~\cite{qin2025tribology}. The resulting friction coefficients can be quantified with high accuracy using, for example, a surface forces apparatus or atomic force microscopy~\cite{hayler2024surface,butt2005force}, thereby enabling the design of protocols targeting specific packing structures. To obtain looser closed structures, the friction coefficient between particles should be increased. Conversely, to achieve minimal moment packings at relatively high evaporation rates, the interparticle friction coefficient must be reduced.

\section{DISCUSSION}

In this study, we systematically investigated the packing dynamics of colloidal clusters driven by emulsion droplet evaporation, with a particular focus on the effects of evaporation rate and interparticle friction for different number of colloidal particles. Numerical simulations demonstrate that diverse packing configurations, including open, closed, and minimal moment structures, can be accessed by tuning the Capillary number ($Ca$) and friction coefficient ($\mu$). 

A regime diagram was constructed to illustrate the relationship between these dynamic parameters and the resulting clusters. For smooth particles, the final packing configuration is determined by the competition between capillary and hydrodynamic forces. With increasing particle number, the boundary between open and closed packings remains the same, but that between closed and minimal moment packing shifts to lower $Ca$.

Introducing particle friction significantly broadens the range of conditions under which closed packing can be achieved, reminiscent of previous experimental observations~\cite{cho2007complex}. For rough particles with $\mu > 0.05$, the transition between minimal energy and closed packing regimes follows a power-law relationship, as given in Equation~\eqref{powerlaw}. This relationship is independent of the number of particles. 

Force analysis reveals that the packing dynamics consist of two distinct stages. In the initial packing stage, the interplay between capillary and hydrodynamic forces dominates and determines the critical intermediate structure. In this stage, the initial distribution determines the final packing when capillary force is not dominant. The larger initial second moment will lead to the generation of non-minimal moment final packing structure. In the subsequent rearrangement stage, frictional interactions play the leading role by dictating whether the system can further evolve into its minimal moment configuration. 

Overall, this work highlights the importance of dynamic parameters in evaporation-driven colloidal packing. Tuning evaporation rate and interparticle friction provides a potential strategy for selectively producing desired clusters while suppressing undesired ones. These insights offer guidance for designing colloidal clusters with tailored structures and functions, which is essential for the bottom-up fabrication of advanced materials. There are numerous avenues for future work. For instance, it is interesting to study the packing dynamics of particle mixtures with controlled size disparity to quantify the impact of system asymmetry and to investigate particles with tunable wettability to broaden the applicability of this framework. Another key direction is to develop a predictive theoretical description of how $Ca$ governs the force balance and the evolution of $\delta_\mathrm{t}$, thereby providing deeper insight into the observed power-law relationship in Equation~\eqref{powerlaw}.

\newpage

\section*{METHODS}


To simulate emulsion-assisted packing of colloidal particles, we employ a lattice Boltzmann (LB) method coupled with a discrete element method (DEM). A phase-field LB model is used to resolve the multiphase flow, while DEM is applied to compute particle dynamics. Below, we briefly summarize the governing equations; additional methodological details are provided in Text~S1, and numerical validation is presented in Text~S2. Further details can be found in our previous work ~\cite{naga2025modelingdropletparticleinteractionssolid}.

\subsection*{Governing Equations for Multiphase Fluid Dynamics}

For the emulsion droplet and surrounding fluid, the fluid flow is described by the incompressible Navier-Stokes (N-S) equations as follows:
\begin{equation}
  \nabla \cdot \mathbf{u} = 0,
  \label{NS-mass}
\end{equation}
\begin{equation}
  \frac{\partial (\rho \mathbf{u})}{\partial t} + \nabla \cdot (\rho \mathbf{u} \mathbf{u}) 
= -\nabla p + \nabla \cdot \left[ \rho \nu \left( \nabla \mathbf{u} + (\nabla \mathbf{u})^T \right) \right] 
+ \mathbf{f}_{\mathrm{c}} + \mathbf{f}_{\mathrm{b}},
  \label{NS-moment}
\end{equation}
where $\mathbf{u}$ is the fluid velocity, $\rho$ is the fluid density, $t$ is time, $p$ is the pressure, $\nu$ is the kinematic viscosity, $\mathbf{f}_{\mathrm{c}}$ is the capillary force at the interface, and $\mathbf{f}_{\mathrm{b}}$ is the body force. Because the particles are confined within the finite volume of an emulsion droplet, thermal fluctuations have a limited influence and are neglected in the present study.

In addition to the Navier–Stokes equations, an order parameter $\phi$ governed by an advection diffusion equation is introduced to distinguish between the two immiscible fluids, A (emulsion droplet) and B (surrounding fluid). In this work, an improved conservative Allen–Cahn (AC) equation is employed, as proposed in~\citep{liang2023lattice}
\begin{equation}
  \frac{\partial \phi}{\partial t} + \nabla \cdot (\phi \mathbf{u}) = \nabla \cdot M \left\{ \nabla \phi - \frac{1}{W} \left[ 1 - \tanh^2 \left( \frac{1}{2} \ln \left( \frac{\phi}{1 - \phi} \right) \right) \right] \mathbf{n} \right\}+\frac{\dot{m}'''}{\rho_\mathrm{A}}
  \label{AC}.
\end{equation}
The order parameter $\phi$ distinguishes the two fluids: $\phi=1$ in fluid~A and $\phi=0$ in fluid~B. The mixture density is interpolated as
\begin{equation}
  \rho = \rho_{\mathrm{B}} + \phi \left( \rho_{\mathrm{A}} - \rho_{\mathrm{B}} \right).
  \label{rho-phi}
\end{equation}
$M$ denotes the mobility, and $W$ is the characteristic thickness of the interface. 
The unit normal vector at the interface is defined as $\mathbf{n} = \nabla \phi / |\nabla \phi|$. $\dot{m}'''$ denotes the evaporation mass source term of fluid A defined by \cite{sugimoto2021consistent}
\begin{equation}
  \dot{m}'''=\rho_\mathrm{A}U_\mathrm{e}\mathbf{n}\cdot \nabla \phi=\rho_\mathrm{A}U_\mathrm{e} |\nabla \phi|,
\end{equation}
where $\rho_\mathrm{A}$ is the density of fluid A. For simplicity, we treat the volumetric evaporation rate $\dot{m}'''$ as a constant. In reality, the evaporation flux on the droplet surface has a non-uniform spatial distribution governed by local heat and mass transfer, which can induce capillary and Marangoni flows and thus more complex convection. Because convective effects are not dominant for the problems considered here, we approximate the evaporation rate as uniform. In future work, we plan to incorporate coupled heat and mass transfer and adopt a more realistic evaporation model. \cite{czelusniak2025analytical, sugimoto2021consistent, yang2023lattice}

The capillary force $\mathbf{f}_c$ is computed based on the order parameter as
\begin{equation}
  \mathbf{f}_c = \left[ 4 \beta \phi (\phi - 1)(\phi - 0.5) - \kappa \nabla^2 \phi \right] \nabla \phi,
  \label{fcAC}
\end{equation}
where $\beta = \frac{12\gamma}{W}$, $\kappa = \frac{3\gamma W}{2}$ and $\gamma$ is the surface tension. By substituting Equations~\eqref{rho-phi} and ~\eqref{fcAC}, the AC equation~\eqref{AC} is consistently coupled to N-S equations~\eqref{NS-mass}-~\eqref{NS-moment}. The above governing equations are solved using the phase-field LB method and the numerical details are provided in Text~S1.

\subsection*{Governing Equations for Particle Dynamics}
The discrete element method is adopted to capture the particle dynamics by calculating the forces on each particle. The capillary force $\textbf{F}_\text{c}$, hydrodynamic force $\textbf{F}_\text{h}$, normal contact force $\textbf{F}_\text{n}$, and friction $\textbf{F}_\text{f}$ and corresponding torque $\textbf{T}$ on each particle are captured and the particle motion and rotation can be determined by Newton’s law \cite{zhang2020new, naga2025modelingdropletparticleinteractionssolid}
\begin{equation}
m_\text{p}\frac{d\mathbf{u}_\text{p}}{dt}=\mathbf{F}=\mathbf{F}_\mathrm{c}+\mathbf{F}_\mathrm{h}+\mathbf{F}_\mathrm{n}+\mathbf{F}_\mathrm{f},
\label{Newton1}
\end{equation}
\begin{equation}
I_\text{p}\frac{d\boldsymbol{\omega}_\text{p}}{dt} = \mathbf{T} = \mathbf{T}_\mathrm{c} + \mathbf{T}_\mathrm{h} + \mathbf{T}_\mathrm{f},
\label{Newton2}
\end{equation}
where $m_\text{p}$ and $I_\text{p}$ are mass and rotational inertia of the particle. $\mathbf{u}_\text{p}$ and $\boldsymbol{\omega}_\text{p}$ are the linear and angular velocities. By computing forces and torques, we integrate particle motion and use the particle positions to update the fluid grids and wall boundary conditions in the LB solver. Details of the force and torque calculations are provided in Text~S1.

\newpage


\section*{RESOURCE AVAILABILITY}


\subsection*{Lead contact}


Requests for further information and resources should be directed to and will be fulfilled by the lead contact, Halim Kusumaatmaja (halim.kusumaatmaja@ed.ac.uk).

\subsection*{Data and code availability}


\begin{itemize}
    \item All original code has been deposited at Zenodo under the DOI 10.5281/zenodo.17913776 and is publicly available as of the date of publication.\cite{junyu_yang_2025_17913776}
    \item Any additional information required to reanalyze the data reported in this paper is available from the lead contact upon request.    
\end{itemize}

\section*{ACKNOWLEDGMENTS}


The research is supported by the UK Engineering and Physical Sciences Research Council (EPSRC) under grants No. EP/V034154/2 (J.Y. and H.K.) and EP/X028410/2 (A.N.), and by the Leverhulme Trust under grant RPG-2022-140 (X.Z. and H.K.). We also acknowledge the Cirrus UK National Tier-2 HPC Service at EPCC funded by the University of Edinburgh and EPSRC (EP/P020267/1), and ARCHER2 supercomputing resources provided by the EPSRC project ‘UK Consortium on Mesoscale Engineering Sciences (UKCOMES)’ (EP/X035875/1).

\section*{AUTHOR CONTRIBUTIONS}


Conceptualization, J.Y. and H.K.; methodology, J.Y., A.N., and X.Z.; investigation, J.Y., A.N., X.Z., and H.K.; writing-–original draft, J.Y. and H.K.; writing-–review \& editing, A.N., X.Z., and H.K.; funding acquisition, H.K.; resources, H.K.; supervision, H.K..

\section*{DECLARATION OF INTERESTS}


The authors declare no competing interests.

\section*{SUPPLEMENTAL INFORMATION INDEX}




\begin{description}
  \item Figures S1-S16 and their legends in a PDF
  \item Text S1. Details of the numerical method
  \item Text S2. Details of the numerical validations
  \item Video S1. Numerical results for different density and viscosity ratios
  \item Video S2. Numerical results for different $Ca$
  \item Video S3. Explanation of the role of friction in packing dynamics
\end{description}

\newpage

\section*{MAIN FIGURE TITLES AND LEGENDS}




\noindent\includegraphics[width=\linewidth]{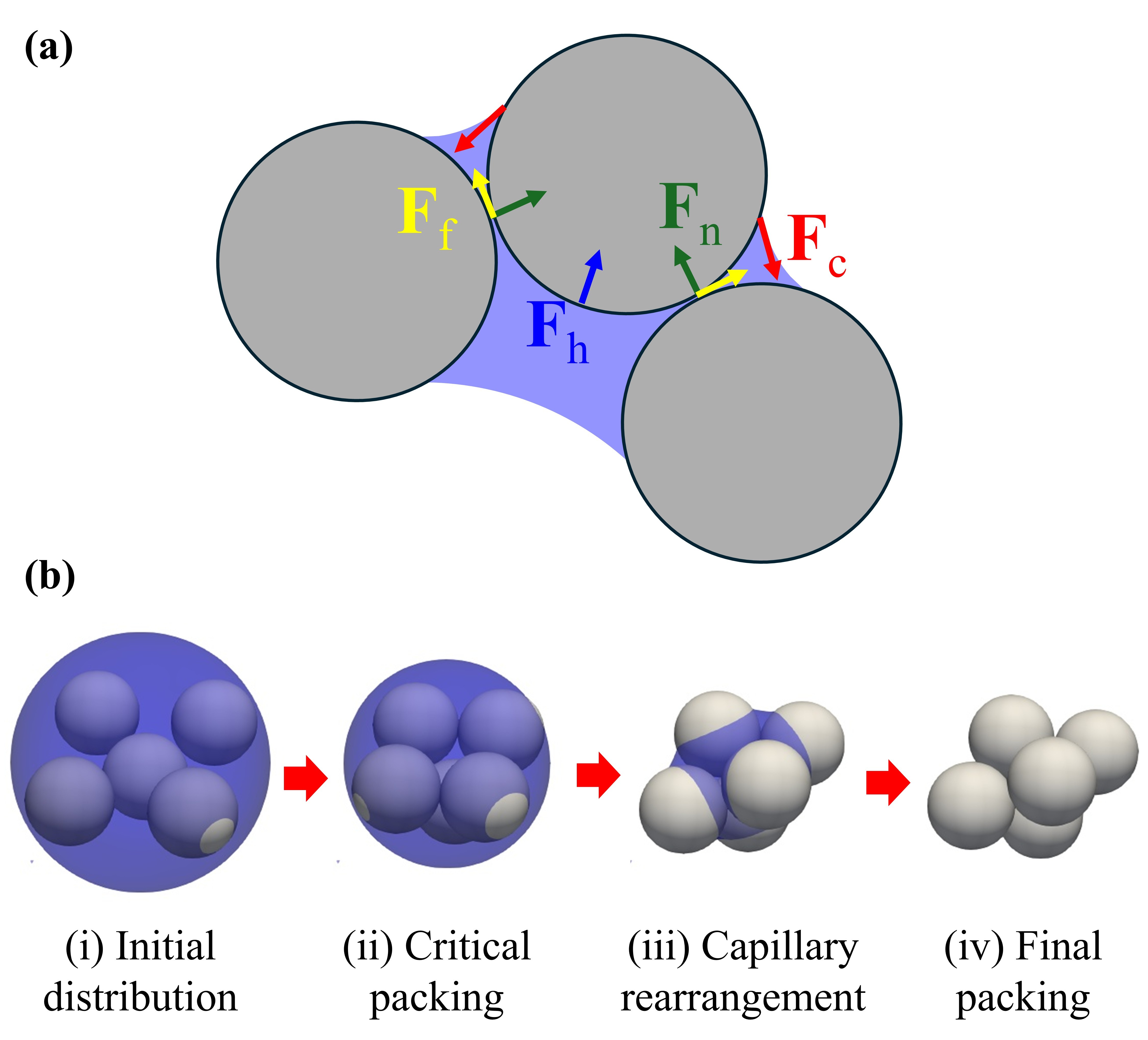}

\subsection*{Figure 1. Schematic illustration of evaporation-driven packing.}

(a) The forces applied in the spherical particles during evaporation-driven packing. $\textbf{F}_\text{c}$ is capillary force, $\textbf{F}_\text{h}$ hydrodynamic force, $\textbf{F}_\text{n}$ contact force, $\textbf{F}_\text{f}$ friction. 
\newline
(b) Schematic representation of spherical particles self-assembly via droplet evaporation. The droplet is represented by blue color and the particle is gray.

\newpage

\noindent\includegraphics[width=\linewidth]{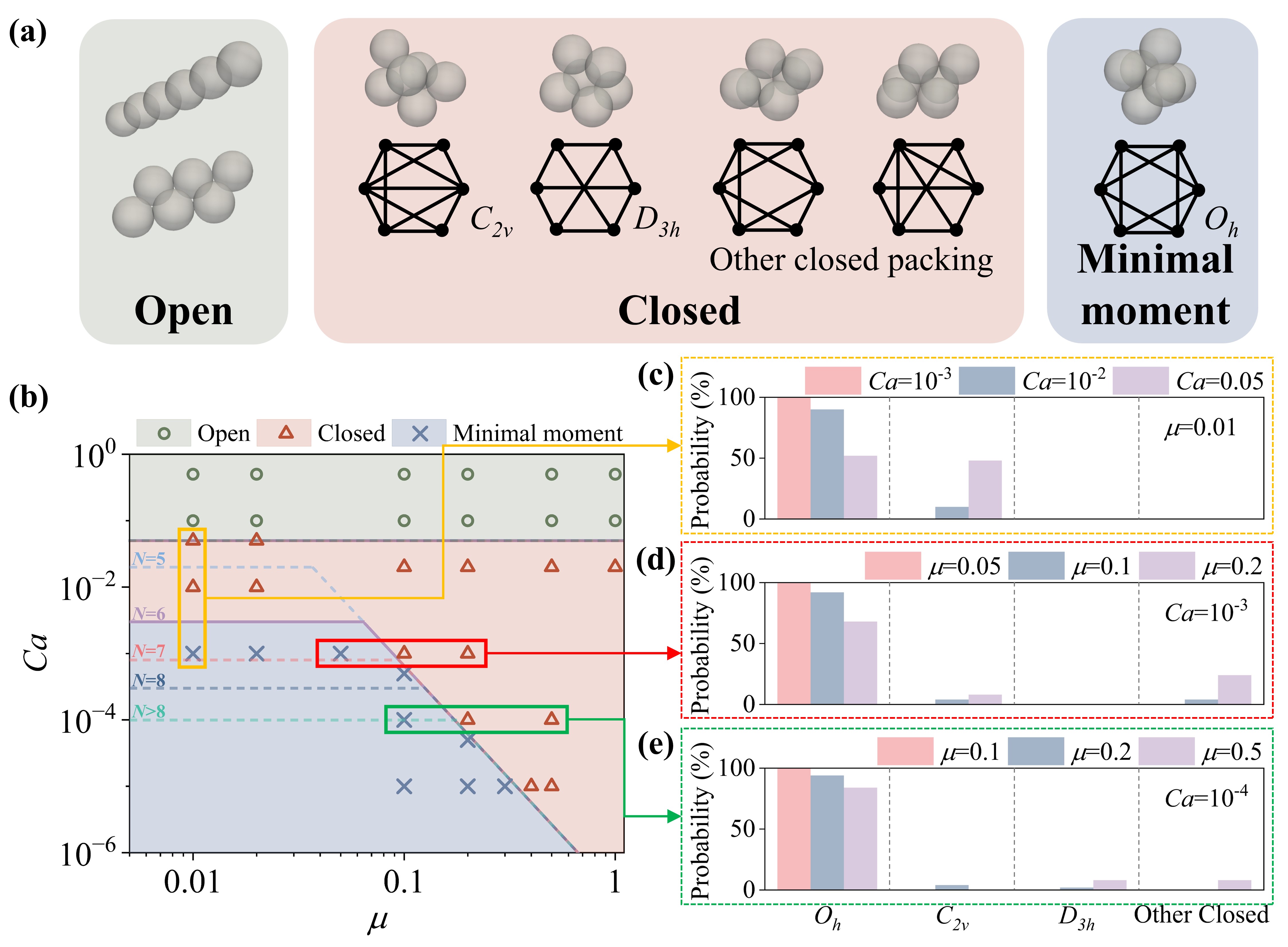}

\subsection*{Figure 2. Packing regime diagram for $N=6$.}

(a) Three typical final packing configurations obtained in simulations. 
\newline
(b) Regime diagram of packing types as a function of capillary number $Ca$ and friction coefficient $\mu$.
\newline
(c)-(e) Probability distributions of distinct packing structures under different conditions highlighted in panel (b).

\newpage

\noindent\includegraphics[width=\linewidth]{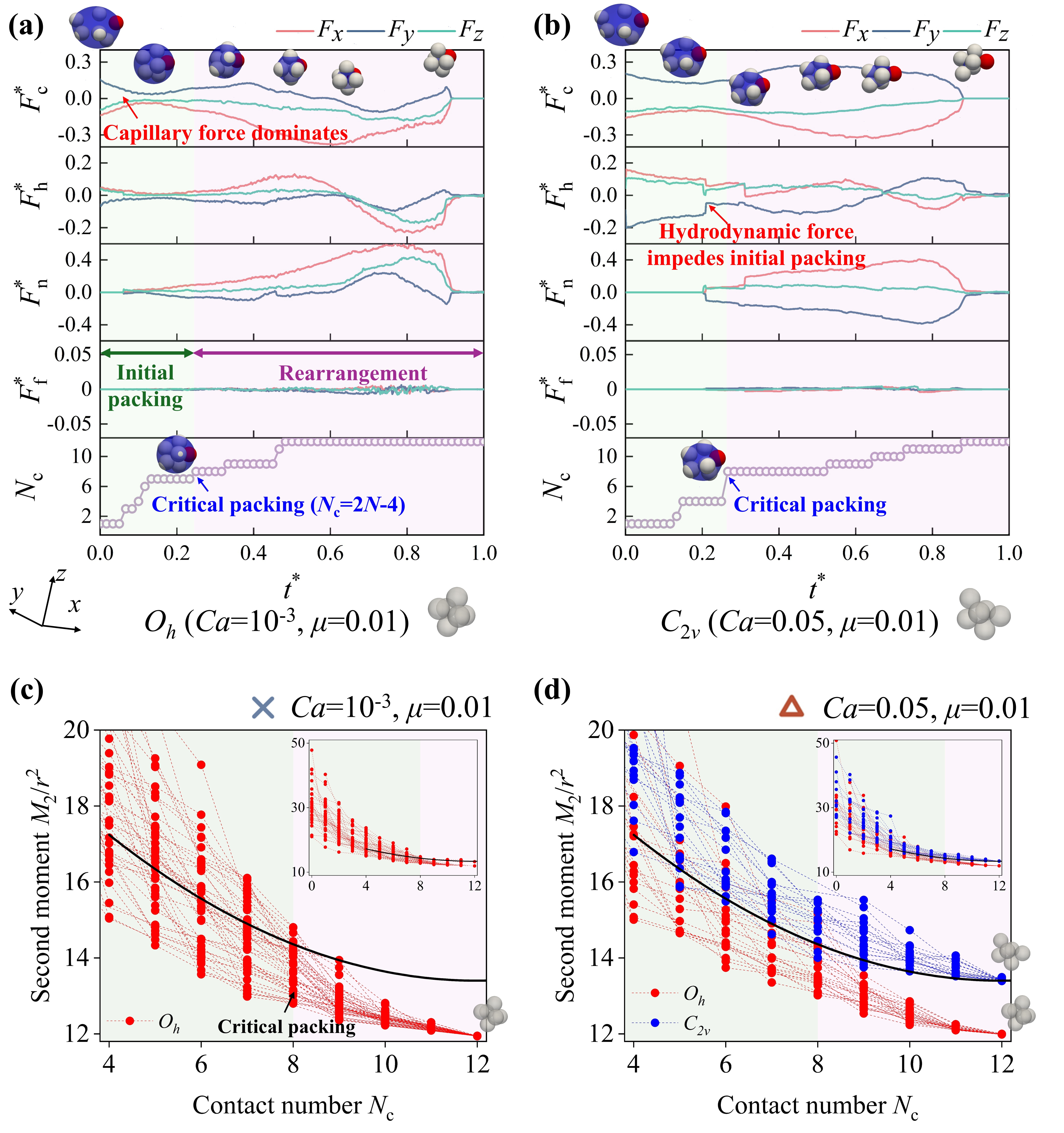}

\subsection*{Figure 3. Force evolution on a selected particle (marked in red) during packing for $N=6$ with smooth particles.}

(a) Dynamic conditions are $Ca=1.0\times10^{-3}$, $\mu=0.01$ (yielding the minimal moment $O_h$ packing); 
\newline
(b) $Ca=0.05$, $\mu=0.01$ (yielding the ground-state $C_{2v}$ packing). The initial particle distributions in (a) and (b) are identical.
\newline
(c, d): $M_2$–$N_\mathrm{c}$ trajectories for $N=6$ under different initial distributions, corresponding to the conditions in (a) and (b), respectively.

\newpage

\noindent\includegraphics[width=\linewidth]{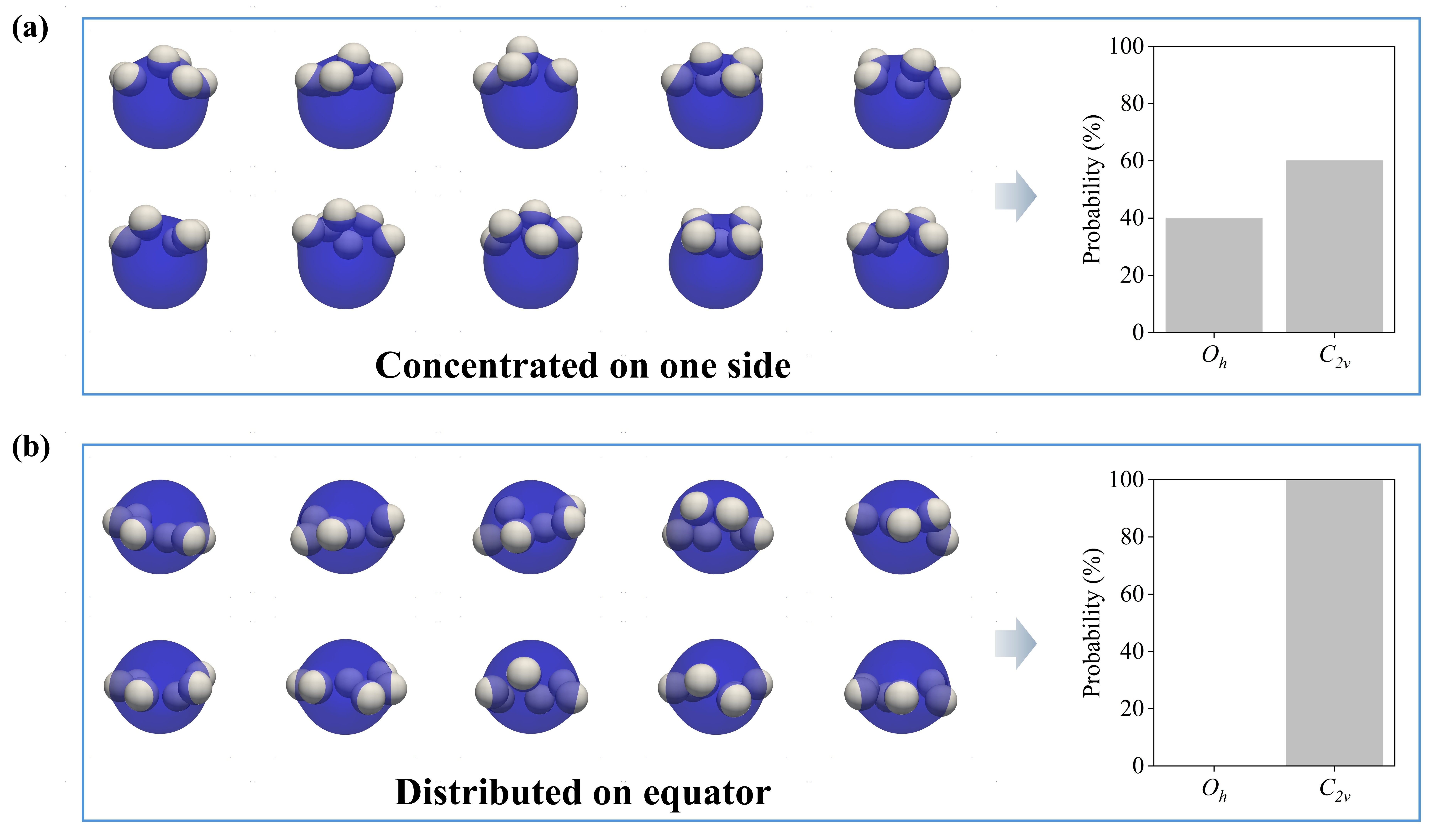}

\subsection*{Figure 4. Representative initial particle distributions for $N=6$ and the corresponding probability of the final packing structures at $Ca = 0.05$ and $\mu = 0.01$.}

(a) Particles concentrated on one side of the droplet: the probability of obtaining the $C_{2v}$ structure is $60\%$. For each type of initial distribution, ten random realizations are simulated. 
\newline
(b) Particles distributed along the equator: the probability of obtaining the $C_{2v}$ structure is $100\%$.

\newpage

\noindent\includegraphics[width=\linewidth]{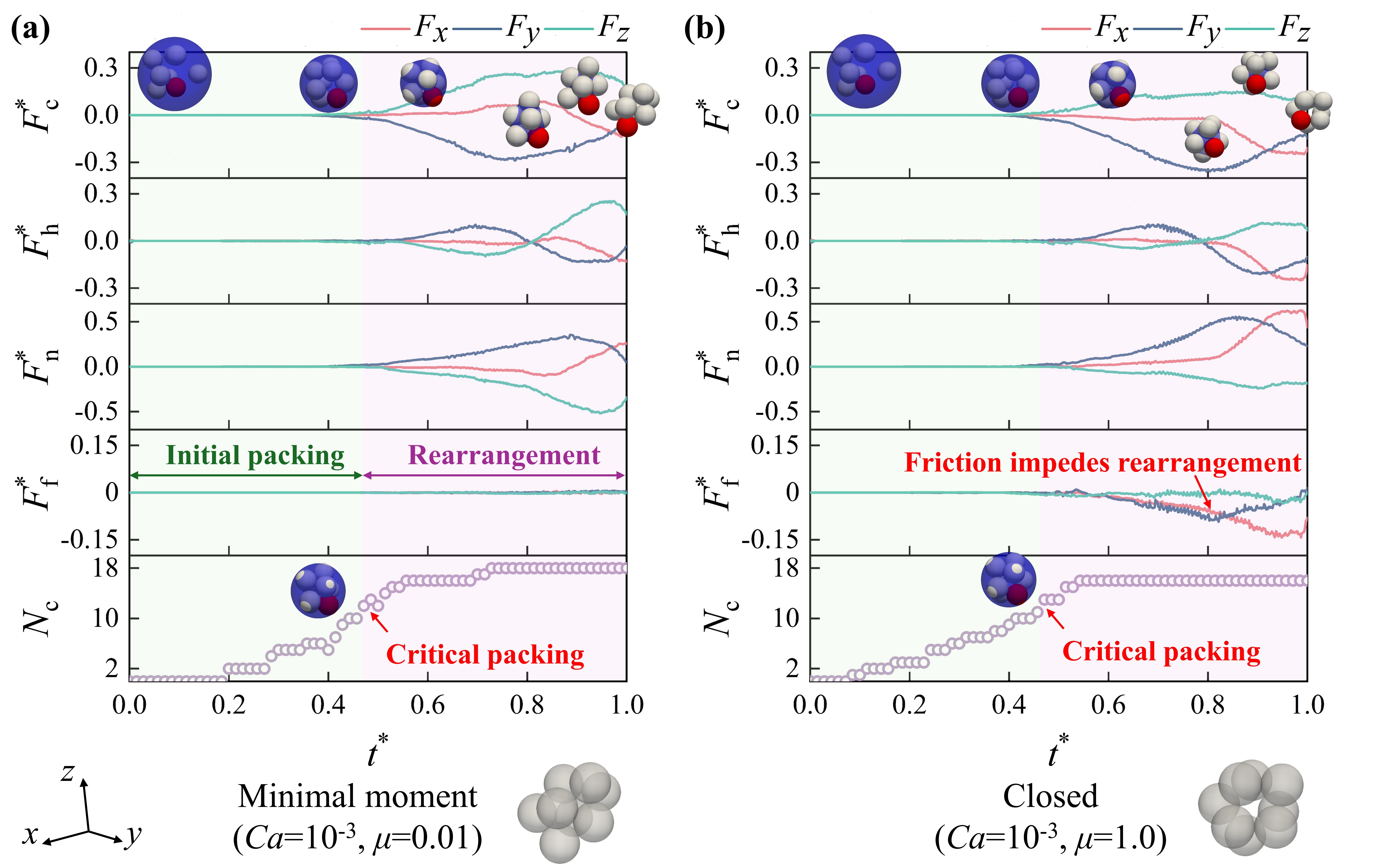}

\subsection*{Figure 5. Force evolution on a selected particle (marked in red) during emulsion-assisted packing for $N=8$.}

The $x$, $y$, and $z$ axes represent spatial directions; subscripts c, h, f, and n denote capillary, hydrodynamic, frictional, and normal contact forces, respectively. The packing dynamic conditions are (a) $Ca = 1.0 \times 10^{-3},\ \mu = 0.01$, (b) $Ca = 1.0 \times 10^{-3},\ \mu = 1.0$, corresponding to minimal moment and closed packing respectively.

\newpage


\bibliography{references}

\bigskip


\end{document}